\documentclass[preprint]{aastex63}
\usepackage{comment}
\usepackage{natbib}
\usepackage{gensymb}
\usepackage{booktabs}
\usepackage{hyperref}

\newcommand\track{}
\received{29 September 2020}
\revised{6 November 2020}
\accepted{7 November 2020}

\submitjournal{AASPSJ}

\shorttitle{Seeing the Bigger Picture: \emph {Rosetta} and Beyond}
\shortauthors{Usher et al.}

\begin{document}

\title{Seeing the Bigger Picture : \\The \track{\emph {Rosetta}} Mission Amateur Observing Campaign and Lessons for the Future}

\correspondingauthor{Helen Usher}
\email{helen.usher@open.ac.uk}

\author[0000-0002-8658-5534]{Helen Usher}
\affiliation{The Open University, Walton Hall, Milton Keynes,
MK7 6AA, UK}

\author[0000-0001-9328-2905]{Colin Snodgrass}
\affiliation{University of Edinburgh, Institute for Astronomy, Royal Observatory Edinburgh, Blackford Hill, Edinburgh, EH9 3HJ, UK}

\author[0000-0002-9153-9786]{Simon F. Green}
\affiliation{The Open University, Walton Hall, Milton Keynes,
MK7 6AA, UK}

\author[0000-0001-7619-8269]{Andrew Norton}
\affiliation{The Open University, Walton Hall, Milton Keynes,
MK7 6AA, UK}

\author[0000-0002-1977-065X]{Paul Roche}
\affiliation{School of Physics and Astronomy, Cardiff University, McKenzie House,
Newport Road, Cardiff, CF24 0DE, UK}

\begin{abstract}
Amateur astronomers can make useful contributions to the study of comets.  They add temporal coverage and multi-scale observations which can aid the study of fast-changing, and large-scale comet features.  We document and review the amateur observing campaign set up to complement the \track{\emph{Rosetta}} space mission, including the data submitted to date, and consider the campaign's effectiveness in the light of experience from previous comet amateur campaigns.  We report the results of surveys of campaign participants, the amateur astronomy community, and schools who participated in a comet 46P observing campaign.  We draw lessons for future campaigns which include the need for: clarity of objectives; recognising the wider impact campaigns can have on increasing science capital; clear, consistent, timely and tailored guidance; easy upload procedures with in-built quality control; and, regular communication, feedback and recognition.
\end{abstract}


\section{Introduction} \label{sec:intro}
\subsection{Background}
Comets are small, active, volatile-rich, solar system bodies. Each comet is unique, but a consistent feature is their unpredictability.  Their appearance can change dramatically over very short timescales:  brightening or fading, rapidly or slowly, breaking apart, exhibiting spectacular tails, or not.  Their constituent parts (nucleus, coma, tails and trails) are on significantly different physical scales: from a nucleus at \textless10 km to tails and trails which can extend many AU.  Comets have a wide range of orbital elements, which can change over time due to gravitational perturbations from solar system bodies, and non-gravitational forces from comet activity (e.g., outgassing). Their position in the sky can change rapidly.  These diverse characteristics make studying comets exciting, but challenging.  Despite being observed and studied over millenia, comets are still not well understood \citep{AHearn2004, meech-review-pre}.  Understanding comet formation and evolution is important in informing and constraining theories of solar system formation and evolution \citep{AHearn2017,aherarn-buildingblocks}.

To observe and characterise comet activity requires observations over many different time periods and intervals, and at different image scales.  Observing a comet over its different apparitions across many years allows its long-term evolution to be monitored. Multiple observations in a single night can pinpoint the start of outburst events, while monitoring over subsequent days and weeks allows morphological changes in the coma, tails, or trails to be analysed.

Between these extremes, short regular observations allow the comet position to be measured, refining its orbit and non-gravitational forces, and monitoring over different time intervals allows analysis of changes due to rotation or seasons to be undertaken.  

As it is impossible to resolve a comet nucleus from Earth, space missions are required for close-up observations.  The Halley missions, \emph{Deep Space 1}, \emph{Stardust} and \emph{Deep Impact} close fly-bys provided snap-shot views of the inner coma and nucleus of five comets. The European Space Agency (ESA) \emph{Rosetta} mission to 67P/Churyumov–Gerasimenko, which orbited the comet and placed the \emph{Philae} lander on its surface, provided the first opportunity to observe surface activity and evolution of a comet for over two years around its perihelion passage. 

These missions have been essential to add a ground-truth element to comet observations, and are resulting in new insights into comet formation and evolution. Each has been supported and complemented by ground-based observing campaigns.

\subsection{Observational Constraints and Opportunities}

It can be difficult for professional observers to cover the wide range of observations needed for analysis of all comets' dynamic features. Professional telescope resources are scarce.  While the proposal method of allocating resources is good for long-term regular monitoring, it can be too rigid when a rapid response is needed to observe short-term changes such as outbursts. Even long-term monitoring is constrained by over-demand for professional facilities. 

The best observing locations are at altitude and away from light pollution, clustered and not longitudinally well spread. This is problematic when comet visibility windows are short, or in periods of bad observing weather.  Large telescopes often cannot image lower than 20-25\degree elevation due to enclosures, and many have a minimum solar elongation for safety, but comets are often brightest and more interesting to study while close to the Sun and often at low altitude. The image scale of large telescopes produces high resolution, but with relatively narrow fields of view. Imaging large-scale features, such as large comae, tails and trails, requires mosaics, taking significant extra telescope time. As a result, cometary science is an area where amateurs can still make important contributions, supplementing observations from professionals. 

Amateurs literally observe for love, being able to choose what, when, where and how to observe. For many, as their interest and expertise deepens they look for more rewarding targets, transient events, longer-term monitoring and/or scientific projects \citep{Bowler2009}. They are free to monitor and observe comets whenever they are visible and weather conditions allow, and can respond quickly to alerts when changes in a comet are noted.  Subject to visibility, multiple images can be taken over a long period during a night,  allowing stacking of data to improve \track{signal-to-noise} ratios and allowing very faint features to be detected.  Amateurs are spread all across the world, which is particularly useful when observability windows are small in any one location due to altitude and hours of darkness.  Good longitudinal coverage allows continuous monitoring for studying rotation effects and transient features. Some amateurs have excellent unobstructed horizons or have mobile equipment and can travel to find suitable observing situations.  Small telescopes can safely image closer to the Sun.  Finally, comets have large-scale features (particularly tails and trails) which are well suited to smaller amateur telescopes with wider fields of view.  

Recently, the greatly reduced cost of high-quality camera technology, telescopes and related equipment, along with sophisticated software, has meant that many more amateur astronomers can now make high-quality, robust observations and undertake complex astrometric, photometric and \track{morphological analyses}.  The growth of internet technologies and social media has meant that it is much easier for amateurs to: access databases such as JPL \track{HORIZONS}\footnote{\url{https://ssd.jpl.nasa.gov/horizons.cgi}} giving accurate ephemerides for planetarium and mount control programs; be alerted to new comets or activity in known comets; share software and techniques; work to consistent standards; share observations in active groups; and upload to data archives.  Additionally, amateurs and students now have real-time access to high-quality, shared telescope facilities (such as iTelescope\footnote{ \url{https://itelescope.net/}}, Las Cumbres Observatory (LCO)\footnote{ \url{https://lco.global/}}, Slooh\footnote{ \url{https://slooh.com/about/about-slooh-education}}, the Open University's PIRATE and COAST \footnote{\url{http://pirate.open.ac.uk/}}, MicroObservatory\footnote{\url{https://mo-www.cfa.harvard.edu/MicroObservatory/}}, and other education-orientated telescope networks).  These facilities are located in favourable locations, \track{ at altitude, chosen for good} observing and weather conditions \track{(much better than most observers' home locations)}, and robust calibration processes.  Robotic scheduling allows for observing even at inconvenient times. 
\\
\\
\subsection{Pro-Am Comet Campaigns}

Amateurs have participated in professionally coordinated observing campaigns in support of space missions including the \emph{Halley Watch} and \emph{Deep Impact/EPOXI} campaigns, as well as for particularly interesting or well-placed comets such as C/2012 S1 ISON and the \lq\lq4*P" Campaign covering the close approaches to Earth of comets 41P/Tuttle-Giacobini-Kresak, 45P/Honda-Mrkos-Pajdusakova, and comet 46P/Wirtanen.

The \emph{Rosetta} mission included a ground-based observation campaign to support and provide context for the in situ activity \citep{Snodgrass2017b}. This campaign included encouraging amateur astronomers across the world to make and submit observations.

There are lessons which can be learnt from a review of the organisation and outputs from these campaigns, to inform future campaigns (e.g., for comet 67P in 2021), and future comet missions such as Comet Interceptor \citep{Snodgrass2019}.

The \emph {Rosetta} amateur campaign has not previously been formally documented and reviewed.  This paper presents details of the data currently available from the campaign.  It documents the results of surveys of \emph {Rosetta} campaign participants, the amateur astronomer community, and some schools who participated in a 46P observing campaign, to inform a discussion on good practice and lessons for future campaigns. \track{While the details may differ, many of the lessons from this campaign are also relevant to other non-comet observing campaigns.}

\section{Previous Comet Campaigns}

\subsection{Halley}
The ground-based \emph{International Halley Watch} campaign in 1986 was a major undertaking, with a budget of \$10 million, thoroughly planned and implemented. The involvement of amateurs was an important element, but was challenging as electronic communication was in its infancy.  Details of positions, requirements, results etc., all needed to be communicated in hard copy.  Observations were made either visually or with film cameras, and the results were posted back to the campaign \citep{Edberg1988, Sekanina1991b,Dunlop2003}. 

The campaign received much publicity and 1,575 people registered, of which 873 submitted observations \citep{Sekanina1991b}.  To ensure consistency, very detailed guidance was provided.  This proved effective, with 90\% of astrometric submissions being used to determine the orbit and so were important for determining the spacecraft's trajectory.  All the observations were published in hard copy, \track{digitized and released on CD in the 90s, and then} made available online\footnote{\url{https//pdssbn.astro.umd.edu/data_sb/missions/ihw/index.shtml}}.  The images have also been subjected to modern filtering techniques to draw out more coma features. 

The final report on the amateur involvement \citep{Sekanina1991b} noted that:
\begin{itemize}
    \item astronomers worldwide contributed useful data;
    \item not all observations made were reported; 
    \item the majority of observers took their efforts seriously enough to submit data; and 
    \item new observers complied with requirements to a greater extent than experienced observers. \lq\lq Do not expect even the most careful and lucid instructions to be followed rigorously.  Even professionals can be wilful on occasion and amateurs additionally lack the insight to appreciate the importance of standardising observing technique.\rq\rq
\end{itemize} 

\subsection{\emph{Deep Impact/EPOXI}}
The \emph{Deep Impact/EPOXI} mission was designed so that most mission-critical science was undertaken from Earth to enable a wider range of observations \citep{AHearn2005}.  A worldwide ground campaign was needed \citep{Meech2005a}.  For the \emph{Deep Impact} stage (9P/Tempel 1), the observations covered the full time-range, from pre-mission characterisation, through impact and post-impact.  A Small Telescope Science Program was established to complement the professional observatories.  For the follow-on mission (\emph{EPOXI}) to 103P/Hartley 2 the amateur data contributed significantly to a multi-wavelength program of near-continuous observations from August 2010 through encounter on 4 November 2010.   The brightness measurements (a key output from amateur data) allowed the development of an ice sublimation model to estimate dust emissions \citep{Meech2005a}.  Initially the program requested amateur measurements based on their observations; later there was a call for submission of raw data sets for further analysis to photometric standards. 

The CARA\footnote{\url{http://cara.uai.it/home}} (Cometary Archive for Amateur Astronomers) group was very active \citep{Milani2007} \track{in observing Tempel 1 around impact}.  Their observations covered nearly every clear night over 10 months, and resulted in 800 photometric observations.  It chose to use the $Af\rho$ measure \citep{Ahearn1984c}, allowing comparison of data from different telescopes, photometric apertures, epochs, and geometrical positions. CARA members used a consistent set of filters (R and I), took many dozens of images per observation date (and calibration frames), and checked quality. They followed a standardised data processing recipe. CARA also provided software to observers to allow them to analyse and calculate the $Af\rho$ value in a consistent way \citep{Milani2007}. The measurements allowed an observation that the $Af\rho$ value increased by 60\% following impact and took two days to return to the previous level.

\subsection{ISON Morphology Campaign}
This 2013 global campaign involved professionals and amateurs, who obtained mostly continuum images to help characterise dust in the coma of comet C/2012 S1 ISON.  ISON was an unusually well-placed and bright comet on a sungrazing orbit, discovered more than a year before its exceptionally close perihelion passage, and consequently well studied over a wide range of wavelengths at professional observatories \citep{Knight2017, Moreno2014}.  The morphology campaign comprised many hundreds of observations made by nearly two dozen groups \citep{Samarasinha2015}.  When at its brightest the comet was only visible for a short period each night due to its small solar elongation.  The distribution of amateur observers across the world meant that good temporal coverage could be achieved.  The data were used to constrain the duration of coma features, look for diurnal changes, constrain grain velocities, and determine the approximate time grains spent in the sunward side of the coma.  The campaign was managed online\footnote{\url{www.psi.edu/ison}}.  Observers were asked to reduce the data before submitting.  The campaign team then enhanced images to look for coma features.  The results were: the data were far from uniform; few observers had access to narrowband filters (used to separate gas and dust signatures in the coma); the low altitude of observing meant high air mass; no features were visible when the comet was at its brightest, but features were seen earlier in the period. While the challenge was to deal with the non-uniformity of the data set, the temporal coverage was of value. 

The overall conclusion for the usefulness of the amateur data was \lq\lq These campaigns may be most valuable in situations where any single observer can only obtain data during a small window of time, but contributions from many such observers...leads to a more complete understanding of the spatial and temporal evolution of the comet.\rq\rq \citep{Samarasinha2015}.

\subsection{4*P Campaign}
The Planetary Science Institute ran the 4*P campaign\footnote{\url{https://www.psi.edu/41P45P46P}}, starting observations in 2017, for comets 41P and 45P and in 2018 for comet 46P. The 46P element was supplemented by a campaign organised by the University of Maryland\footnote{\url{https://wirtanen.astro.umd.edu}}. For 46P, 18 amateur observers submitted observations. These campaigns comprised both professional and amateur observations.

\subsection{Other Comet Campaigns}
In addition to formal campaigns organised by professional astronomers, there are `informal' campaigns that are self-organised within the active amateur comet observing community whenever a particularly bright or interesting comet appears. It has often been the case that such monitoring discovers interesting behaviour and triggers observations by professionals with access to larger facilities, e.g., in the case of a major outburst, such as that of comet 17P/Holmes in 2007 \citep{miles2010}. 

A recent example of an informal amateur-led campaign of observations was that for C/2019 Y4 ATLAS.  The comet brightened significantly through the early part of 2020, with predictions for possible naked-eye visibility.  It was well placed for observing for large parts of the night from northern latitudes, placed close to the zenith, and its appearance coincided with good weather, and the COVID-19 lockdown.  Multiple observers across the world monitored its development, sharing their observations and analysis primarily via a simple comet mailing list and some Facebook groups (notably Comet Watch).  Observations were submitted to Comet Observations Database (COBS)\footnote{\url{https://www.cobs.si/}}, International Comet Quarterly (ICQ)\footnote{\url{http://www.icq.eps.harvard.edu/}}, Minor Planet Center (MPC)\footnote{\url{https://www.minorplanetcenter.net/}} and the British Astronomical Association (BAA)\footnote{\url{https://britastro.org/cometobs/}} (and other) archives.  The comet became very interesting on 19 March 2020 when it started to fragment. Professional astronomers were alerted to the dramatic changes and were successful in applying for \emph{Hubble} observations\footnote{\url{http://tiny.cc/HubbleCometAtlas}}.  There is now a rich, high-cadence archive available for detailed analysis: 740 and 789 observations in BAA and COBS archives (at 2020-8-18) respectively (note overlap of datasets).
\
\\
\section{\emph{Rosetta} Campaign}

The most ambitious comet mission to date is ESA's \emph{Rosetta} mission\footnote{\url{https://www.esa.int/Science_Exploration/Space_Science/Rosetta}} to comet 67P/Churyumov–Gerasimenko, with aims to contribute to the study of comet and solar system origins, and the relationship between cometary and interstellar material.    
 
It was the first long-term mission to orbit, land on, and \lq live with' a comet, making multi-instrument observations \track{for over 2 years}. 
The orbiter instruments included remote sensors (such as cameras and radio receivers) and direct sensors (such as dust and particle analysers) \citep{Glassmeier2007}. The orbiter's cameras made observations of the comet from distances ranging from 672 million km (when waking from hibernation) to just 2.7 km at closest orbit (additionally it observed whilst descending to the comet's surface for its \lq hard landing').  Larger orbits (e.g., at 1500 km) were used to study the plasma environment and the wider coma.  At perihelion the orbiter was at a distance of $\sim$300 km.   
\\

\subsection{67P Ground-based Campaign Logistics}
A ground-based campaign was part of the mission, including both professional and amateur observations, and coordinated with planning of spacecraft operations \citep{Snodgrass2017b}. The ground- and space-based observations combined to serve three key purposes:
\begin{itemize}
    \item monitoring the overall behaviour and activity of the comet in support of the mission;
    \item providing a basis for multi-scale studies – e.g., how does the composition of the coma vary from 10 to 10\hspace{0.1cm}000 km from the nucleus? What are the chemical reactions behind this variation?
    \item allowing comparison between 67P and other comets, and therefore application of the \emph {Rosetta} results to the larger population.
\end{itemize}
Unfortunately, during the active phase of the mission at the comet (January 2014 -- September 2016), 67P was not very favourably positioned for Earth-based observations.  The next apparition, with perihelion in November 2021, is much more favourable. Observations with large professional telescopes were possible from late February 2014 until shortly after the \emph{Philae} landing in November 2014 \citep{Snodgrass2016b}, after which the comet was at low solar elongation for many months. The comet passed through perihelion in August 2015 and was reasonably well placed for observations during the second half of 2015 and the first half of 2016.

The amateur campaign organisation was funded by JPL as part of the NASA contribution to the ESA-led mission. A website was established by JPL to hold the main campaign information.  This was a static site, with real-time interactions taking place via a Facebook group PACA\_Rosetta67P\footnote{\url{https://www.facebook.com/groups/paca.rosetta67p}} (launched in January 2014 and archived in November 2019).  This was used for communication, sharing guidance, discussions and sharing images.  When it was archived it had 203 members.  

The amateur campaign was formally launched in April 2015, following approximately one year of preparation work in parallel with the early part of the \emph {Rosetta} mission (when the comet was still too far from the Sun to be observable by most amateurs), but initial plans to include amateur astronomers were already discussed as early as 2011, at the beginning of coordination efforts for professional observations.  The invitation to contribute stated that \lq All formats of data will be acceptable and encouraged.  ... CCD, DSLR images, spectra, sketches, visible observations.  ...most helpful will be raw, unprocessed and in FITS format\rq.  Further, more detailed, guidance was issued on 5 June 2015 with guidelines on what observations were required, including filters, orientation and format. \track{On filters, \lq at a minimum, continuum images (UBVRI), but LRGB, or specific narrow band filters (eg OIII) are also acceptable, for studying colours of the comet. We recommend Sloan r’ and g’ filters for a consistent set of data on dust and gas.\rq } It was stated that submissions should include unenhanced images (targets, darks and flats, if any).  The need for accurate time information was stressed.

Each observer was asked to complete a user agreement form, which collected contact details and some basic information on the telescope(s) to which they had access. The data format and filename requirements were set out in detail, along with a request for supplementary information regarding the observations (context information including date/time, location, camera, filter, exposure times, position angle, plate scale -- but not telescope details).
Of the 327 people who registered, 26 FITS format data sets (from individuals or collaborations) are known to have been submitted.  This is a relatively low number, and it is likely that more amateurs hold observations of comet 67P which could be usefully added to the data set for the next analytical stage of this research.  Observers are encouraged to contact the lead author if they wish to contribute observations.

The ESA/Planetary Science Archive (PSA) set up registered user accounts for FTP upload, which were used by some observers. Although this was intended to be the single route for all data collection, delays in setting it up (not available until late September 2015) and initially a lack of clear instructions and/or assistance in using the FTP protocol meant that most users did not use it (a campaign member later documented the process for her fellow observers). Apparent confusion between the requirements for this temporary collection FTP site and the more complicated rules for permanently archived data at the PSA also appeared to put off some users. The JPL project manager set up a Dropbox alternative, and most users submitted this way.  These files were renamed to a standard file-naming convention, which included the date and time of observation, filter, exposure length and initials of observer. The intention was for a subset of the observations to be permanently archived and made publicly available, following some quality assessment. When funding ceased, the personnel involved moved on to other projects, and this meant work on collation and archiving effectively ceased.  At the time of writing there are still two separate locations holding data (with overlap).  The PSA data have not been renamed to match the JPL conventions.  Table 1 contains data from both repositories.

As well as the science data in FITS format uploaded to servers, other images and observations were uploaded to the Flickr\footnote{\url{https://www.flickr.com/groups/paca_67p/}} and/or Facebook\footnote{\url{https://www.facebook.com/groups/paca.rosetta67p}} 67P PACA groups.  Certificates of appreciation were made available to those who took part in the Facebook group, and these were well received.

\subsection{Data submitted}

\subsubsection{FITS Data}

With so many different observers, using such a wide range of equipment and workflows, and of different experience levels, it is inevitable that the data set and the associated metadata varies widely in quantity and quality.  It is not always clear whether/how observations have been calibrated/reduced.  The lack of robust metadata was potentially particularly problematic for detailed analysis - filter and sensor details in particular.

Given the relatively small number of observers, it has been possible to contact most observers and ask for data and FITS header information to be verified and supplemented (subsection \ref{regsurvey}). It has not been possible to reach all observers though as some email addresses appear to be no longer valid, and contact details are not available for those who did not register initially.

\begin{figure}[hbt!]
\centering
\includegraphics[width=16cm]{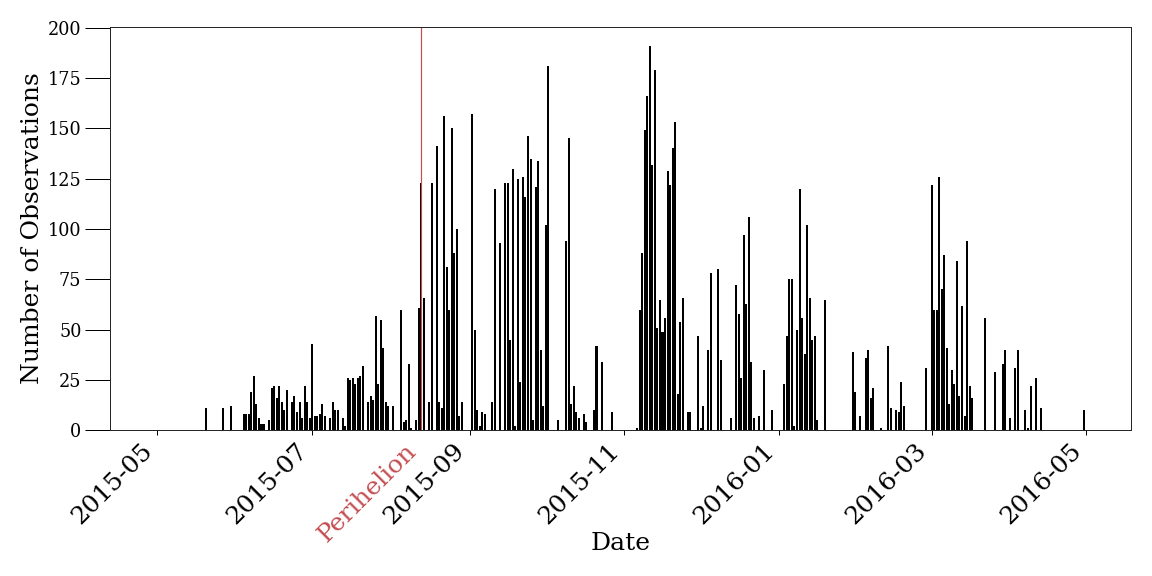}
\caption{Number of Amateur Observations 2015-2016}
\label{fig:obsdate_15-16}
\end{figure}

An analysis of the data set (Table \ref{tab:obs}) shows:
\begin{itemize}
    \item 10,432 observation files known to have been submitted by 26 observers/observing groups covering 284 dates (48 dates and 308 observations were during the previous perihelion passage in 2008-2009).  Figure \ref{fig:obsdate_15-16} shows observations over the main 2015-2016 observing period;
   \end{itemize}

\startlongtable

\begin{longrotatetable}

\begin{deluxetable}{p{6.2em}p{6.43em}cclp{6.em}p{3.355em}p{3em}p{3em}p{5.9em}p{3em}}




\tablecaption{Observations Submitted by Amateur Astronomers - FITS Format}

\tablenum{1}
\tablehead{\colhead{Observers} & \colhead{Period} & \colhead{Dates} & \colhead{Obs} & \colhead{Filters} & \colhead{Locations} & \colhead{Obs} & \colhead{Ap\tablenotemark{b}} & \colhead{FL\tablenotemark{c}} & \colhead{FOV\tablenotemark{d}} &\colhead{Scale}\\ 
\colhead{ } & \colhead{ } & \colhead{(no)} & \colhead{(no)} & \colhead{} & \colhead{} & \colhead{Code\tablenotemark{a}} & \colhead{(mm)} & \colhead{(mm)} & \colhead{ (arcmin)} &\colhead{(arcsec/pix)}\\ } 

\startdata
T Angel and C Harlingten  & 2014-07-08 to 2016-03-15 & 89    & 5659  & C     & Spain & Z85   & 100   & 400   & 11.7x7.79 & 0.92 \\[1.7ex]
J Loum & 2015-10-21 to 2016-04-04 & 40    & 1545  & C,Q,R,G & USA   & W14   & 254   & 1194  & 25.2x18.8 & 1.10 \\[1.8ex]
Slooh\tablenotemark{e} & 2014-11-12 to 2015-07-31 & 46    & 645   & L,R,G,B & Tenerife, Chile & G40, W88 & 350, 355, 432 & 3850, 3904, 2929 & 12.8x8.61, 31.3x20.8, 43.4x43.4 & 0.70, 1.41, 1.91 \\[1.8ex]
E Bryssinck\tablenotemark{e}, F-J Hambsch & 2014-06-27 to 2016-01-15 & 81    & 469   & C,  R & Chile, Belgium, Australia & G39, B96, Q62 & 400, 400, 700 & 2700, 1520, 4531 & 47.0x47.0, 36.5x54.7, 27.8x27.8 & 1.38, 0.82, 1.09 \\[1.8ex]
N Hidenori & 2015-11-30 to 2016-04-09 & 21    & 423   & L     & Japan & Q21   & 400   & 1520  & 84.0x84.0 & 1.23 \\[1.8ex]
F Garcia & 2008-8-23 to  2016-04-03 & 52    & 329   & Clear & Spain & J38   & 250   & 2030  & 17.1x17.1 & 2.00 \\[1.8ex]
P Carson & 2105-07-18 to 2016-04-30 & 34    & 324   & L     & UK    & K02   & 315   & 1656  & 36.7x27.7 & 1.32 \\[1.8ex]
A Chapman & 2016-01-09 to 2016-03-13 & 4     & 245   & r'    & Argentina &   & 203   & 807   & 38.2x28.6 & 1.65 \\[1.8ex]
A Diepvens & 2015-08-11 to 2016-01-11 & 17    & 207   & L,R   & Belgium & C23   & 200   & 1350  & 34.5x23.2 & 1.89 \\[1.8ex]
\tablebreak
M Tsumara\tablenotemark{f} & 2015-07-14 to 2016-04-11 & 13    & 120   & C,R,G,B & ?     & ?     & ? & ? & 249.0x165.6 & 3.72 \\[1.8ex]
P Lake\tablenotemark{e} & 2014-03-07 to 2015-08-08 & 5     & 88    & L,V,R,I & USA   & H06   & 510, 508 & 2260, 2280 & 55.7x55.7, 54.7x36.5 & 1.09, 0.82 \\[1.8ex]
W Clark\tablenotemark{e} & 2015-08-13 to 2015-12-02 & 6     & 60    & L,R,G,B & USA   & H06, U69, H06 & 508, 610, 430 & 2280, 3962, 1939 & 32.1x32.1, 32.1x32.1, 49.0x32.7  & 0.63, 1.26, 0.96 \\[1.8ex]
R Castillo & 2015-08-13 to 2016-03-13 & 3     & 54    & L     & Spain &  & 254   & 1194  & 39.6x26.4 & 3.10 \\[1.8ex]
J-P Nougayrede, G Arlic, and F Metz, and C Andre,  & 2016-03-01 & 1     & 52    & L     & France & 586 & 600   & 2002  & 47.5x31.7 & 0.93 \\[1.8ex]
Northolt Branch Obs & 2016-01-15 to 2016-03-05 & 3     & 51    & None  & UK    &  Z80 & 71    & 418   & 73.3x54.7 & 3.16 \\[1.8ex]
N Howes\tablenotemark{e} & 2012-04-25 to 2013-07-05 & 3     & 41    & R     & Australia, USA & E10, F65 & 2000  & 20000 & 10.2x10.2 & 0.3 \\[1.8ex]
T Traub\tablenotemark{f} & 2014-7-22 to 2106-03-29 & 8     & 26    & L,R,G,B & USA   &  & 610   & 7788  & 16.4x16.4 & 0.32 \\[1.8ex]
J Chambo\tablenotemark{e} & 2015-06-24 to 2015-11-19 & 4     & 17    & L,R,G,B & Australia, USA, \, \,\,\, \,\,USA & Q62, H06, U69 & 510, 508, 610 & 2260, 2280, 3962 & 32.1x32.1,  32.1x32.1, 32.1x32.1&0.63, 0.63, 1.26 \\[1.8ex]
\tablebreak
J L Maestre\tablenotemark{f} & 2015-11-18 & 1     & 15    &       & Spain &   & 406   & 3900  & 21.7x21.7 & 1.27 \\[1.8ex]
P Brlas\tablenotemark{e} & 2014-06-17 to 2015-12-19 & 12    & 14    & L,V,R & Australia, Australia, USA, \, \,\, Spain, \, \,\, USA & Q62, Q62, U69, I89, H06 & 700, 430, 610, 318, 106 & 4527, 2912, 3962, 2541, 530 & 27.8x27.8, 43.9x43.9, 32.1x32.1, 37.3x24.9, 234x156 & 0.55, 0.64, 0.63, 0.73, 3.51 \\[1.8ex]
M Tissington  (SARAS) & 2015-10-17 to 2016-01-10 & 5     & 13    & C     & Tenerife & J54   & 355   & 1877  & 24.4x24.4 & 1.43 \\[1.7ex]
R Nicollerat & 2015-10-10 & 1     & 12    & C     & Switzerland & K17   & 354   & 2937  & 13.5x13.6 & 0.79 \\[1.8ex]
T Zwach\tablenotemark{f} & 2016-04-07 to 2016-04-08 & 1     & 11    & L,R,V,B & Spain & I89   & 150   & 1095  & 111.6x74.4 & 1.67 \\[1.8ex]
K Yoshimoto & 2015-07-25 to 2016-03-07 & 5     & 9     & C,V,I & Japan &   & 160   & 1000  & 35.2x35.2 & 4.1 \\[1.8ex]
Isle of Man Observatory & 2016-02-18 to  2016-03-15 & 2     & 2     & None  & Isle of Man & 987 & 406   & 4064  &  9.0x9.0 & 1.06 \\[1.8ex]
P Detterline & 2016-01-06 & 1     & 1     & C     & Australia &  & 356   & 1914  & 24.1x16.2 & 0.66 \\[1.8ex]
\enddata
\tablecomments{The table is sorted by the number of images submitted by each observer.} 
\tablenotetext{a}{MPC/IAU Observatory code if applicable.}
\tablenotetext{b}{Telescope Aperture}
\tablenotetext{c}{Telescope Focal Length}
\tablenotetext{d}{Image Field of View}
\tablenotetext{e}{These observers used multiple telescope configurations either locally or remotely.}
\tablenotetext{f}{It has not yet been possible to confirm these data with the observer.}




\label{tab:obs}
\end{deluxetable}
\end{longrotatetable}

\begin{figure}[hbt!]
\centering
\includegraphics[width=16cm]{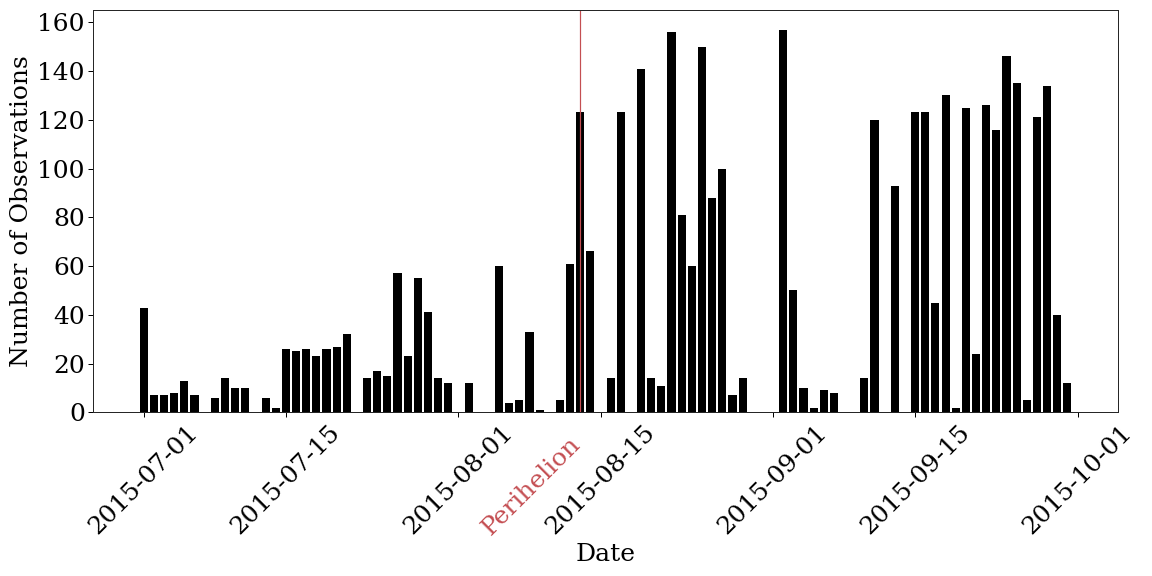}
\caption{Number of Amateur Observations Around Perihelion (13 August 2015)}
\label{fig:obsdate-peri}
\end{figure}

\begin{itemize}
    \item there is good geographical coverage (Figure \ref{fig:obslocation});
    \item there is good temporal coverage around perihelion on 13 August 2015, and around the dates of particular interest identified so far when outbursts were noted by spacecraft instruments in July, August and September 2015 \citep{Vincent2016} (Figure \ref{fig:obsdate-peri});
    \item there are wide ranges of apertures, fields of view, and pixel scales used for observations (Table 1);
     \item some observers made just a small number of observations each night, others acquired multiple images in different filters;
    \item Tony Angel and Caisey Harlingten’s data set is by far the largest in number, with a large number of images per night;
    \item only 8 observers provided calibration/reduction files (578 files) as requested in the guidance, although others submitted calibrated images.  Some submitted stacked images rather than unprocessed images;
    \item 993 observations were undertaken with remote telescopes, which have standard pipeline calibration processes;
    \item the information in the FITS header does not always conform to the guidance or to FITS standards;
        \begin{figure}[hbt!]
\centering
\includegraphics[width=16cm]{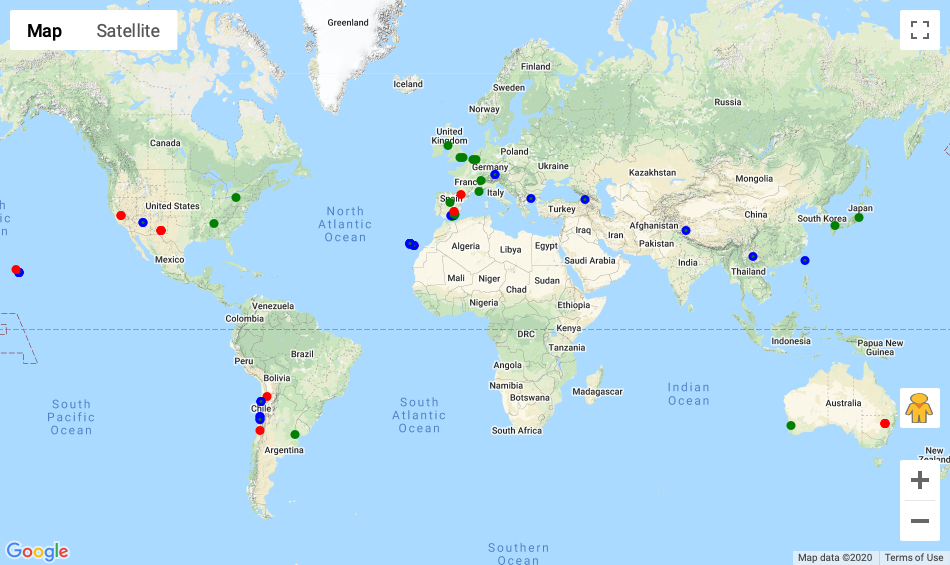}
\centering
\caption{Observing locations (Blue=Professional, Red=Remote, Green=Local)} 
\label{fig:obslocation}
\end{figure}
    \item a variety of filters have been used, \track{but primarily standard imaging filters ( Clear, Luminance, Red, Green, Blue) rather than scientific filters (UVBRI or Sloan r', g')}. In some cases there is no filter data in the FITS header and so follow up with observers has been needed before analysis;
    \item the guidance asked for a narrative file providing extra details of the observations, but these were not generally provided. For some observers, who did not initially register, this has meant no contact details are available either.
\end{itemize}

\begin{figure}[hbt!]
\centering
\includegraphics[width=18cm]{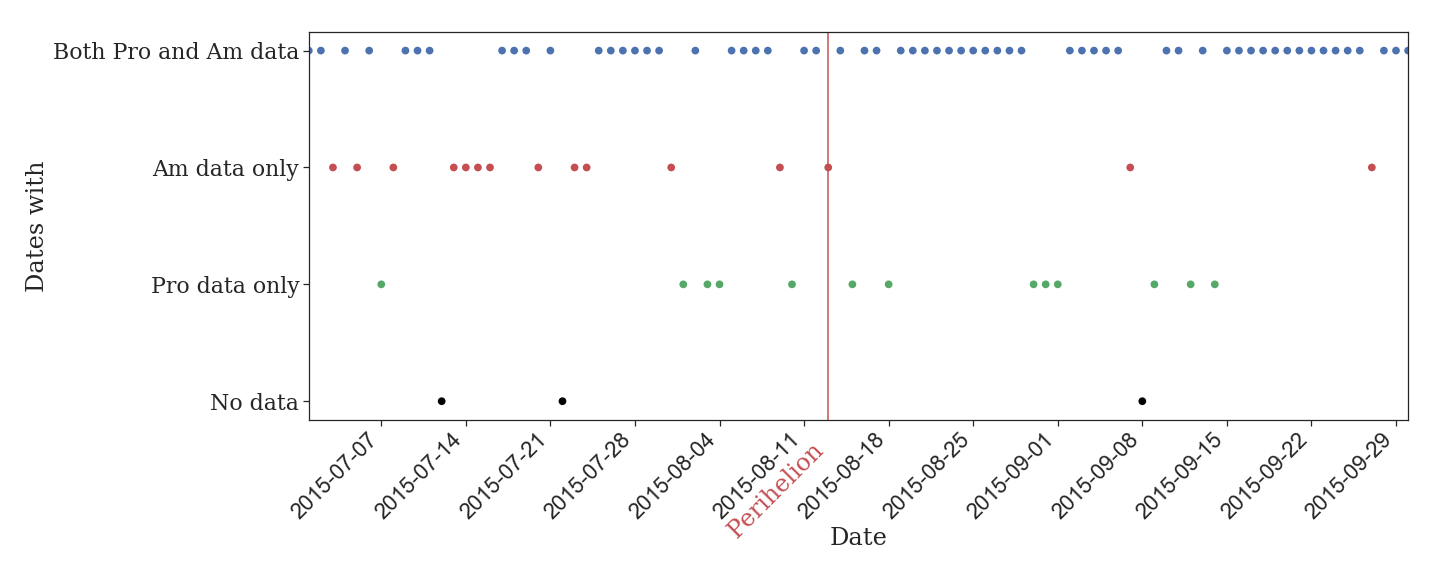}
\centering
\caption{Professional and Amateur Observations Around Perihelion (13 August 2015). The figure shows how amateur observations supplemented the professional observations, with 15 days during the period 2015-07-01 to 2015-09-30, around perihelion, when only amateur observations are available.}
\label{fig:combined-peri}
\end{figure}

Details of the professional observations submitted to the \emph {Rosetta} campaign were obtained.  A comparative analysis of the dates of observations was undertaken. This showed that there were 58 days, during the period 2013-04-17 to 2016-04-30, when amateur observations were available but no professional data were available.  In the 3 months around perihelion (2015-07-01 to 2015-10-01) there were 15 days when only amateur observations were available (Figure \ref{fig:combined-peri}).  The aim of using amateur observations to improve temporal coverage has been achieved.

\subsubsection{Images}
\begin{figure}[hbt!]
\centering
\includegraphics[width=7.5cm]{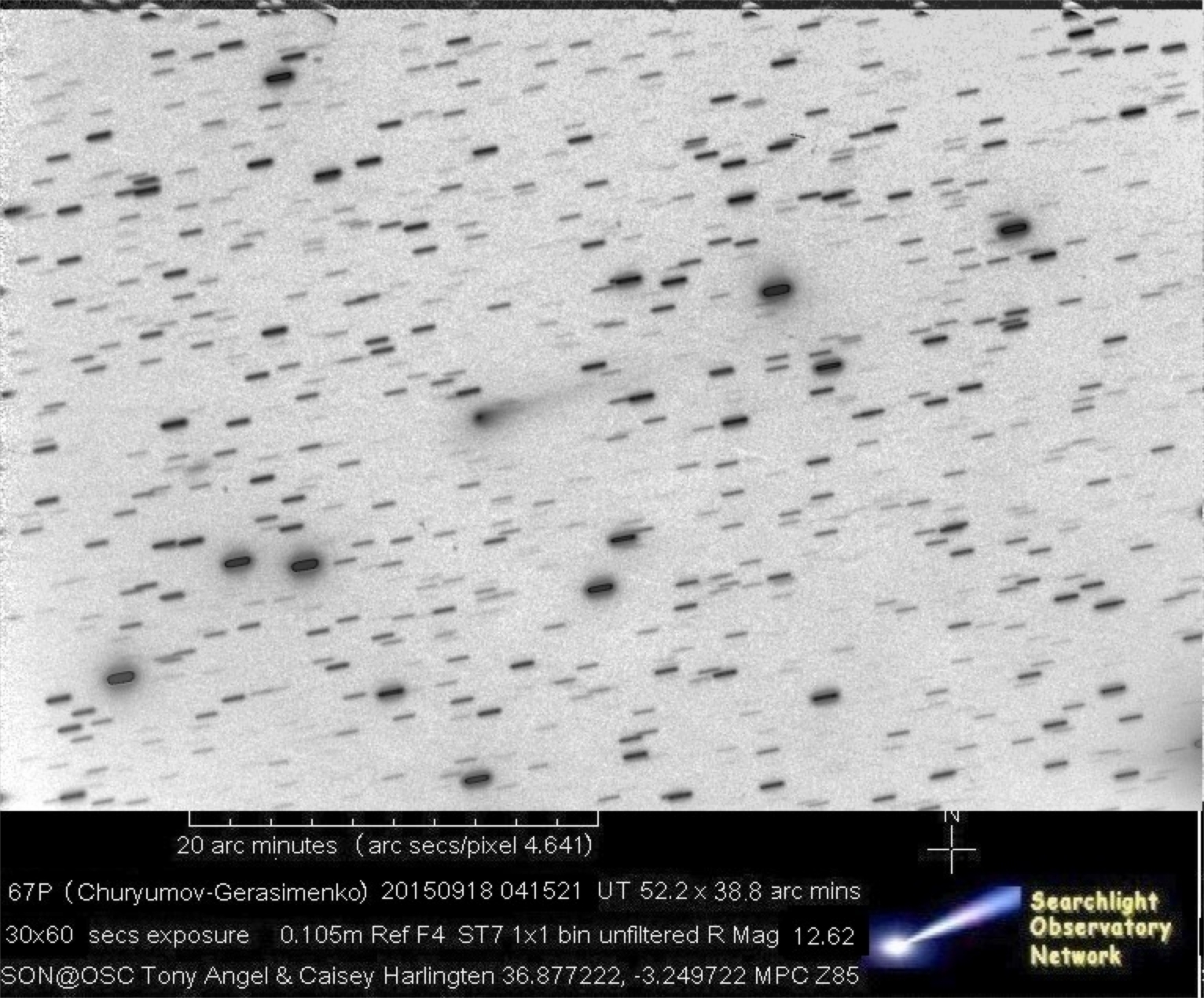}
\centering
\caption{Example of Flickr Upload - scientific analysis of observation on 18 September 2015, wide field showing tail.  Credit: Tony Angel and Caisey Harlingten.}
\label{fig:tonyflickr}
\end{figure}

\begin{figure}[hbt!]

\centering
\includegraphics[width=7.5cm]{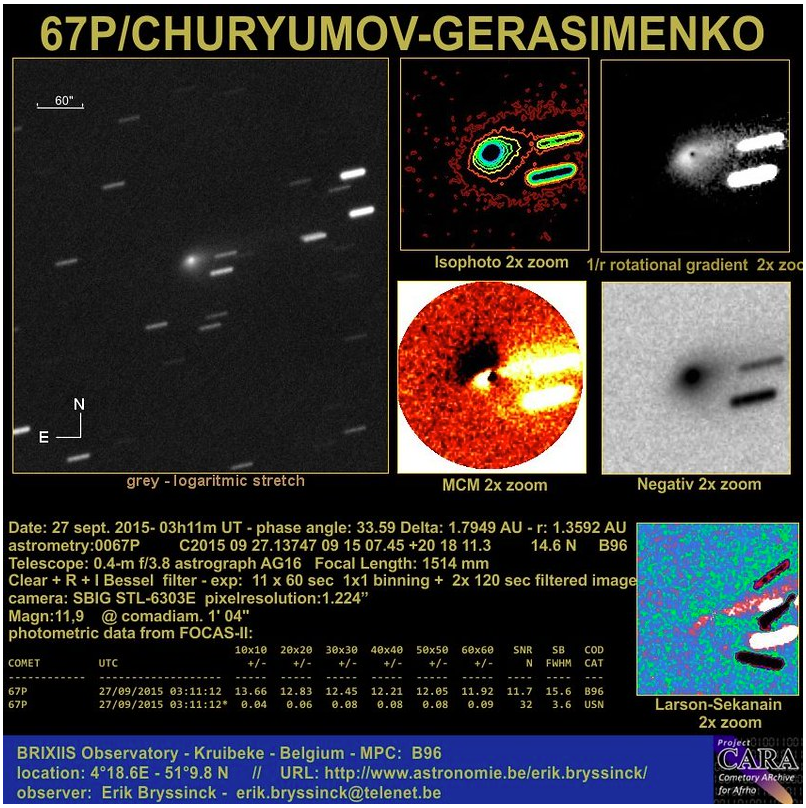}
\caption{Example of Flickr Upload - scientific analysis of observation of coma on 18 September 2015, including coma morphology.  Credit: Erik Bryssinck.}
\label{fig:ebflickr}
\end{figure}

\begin{figure}[hbt!]
\centering
\includegraphics[width=15cm]{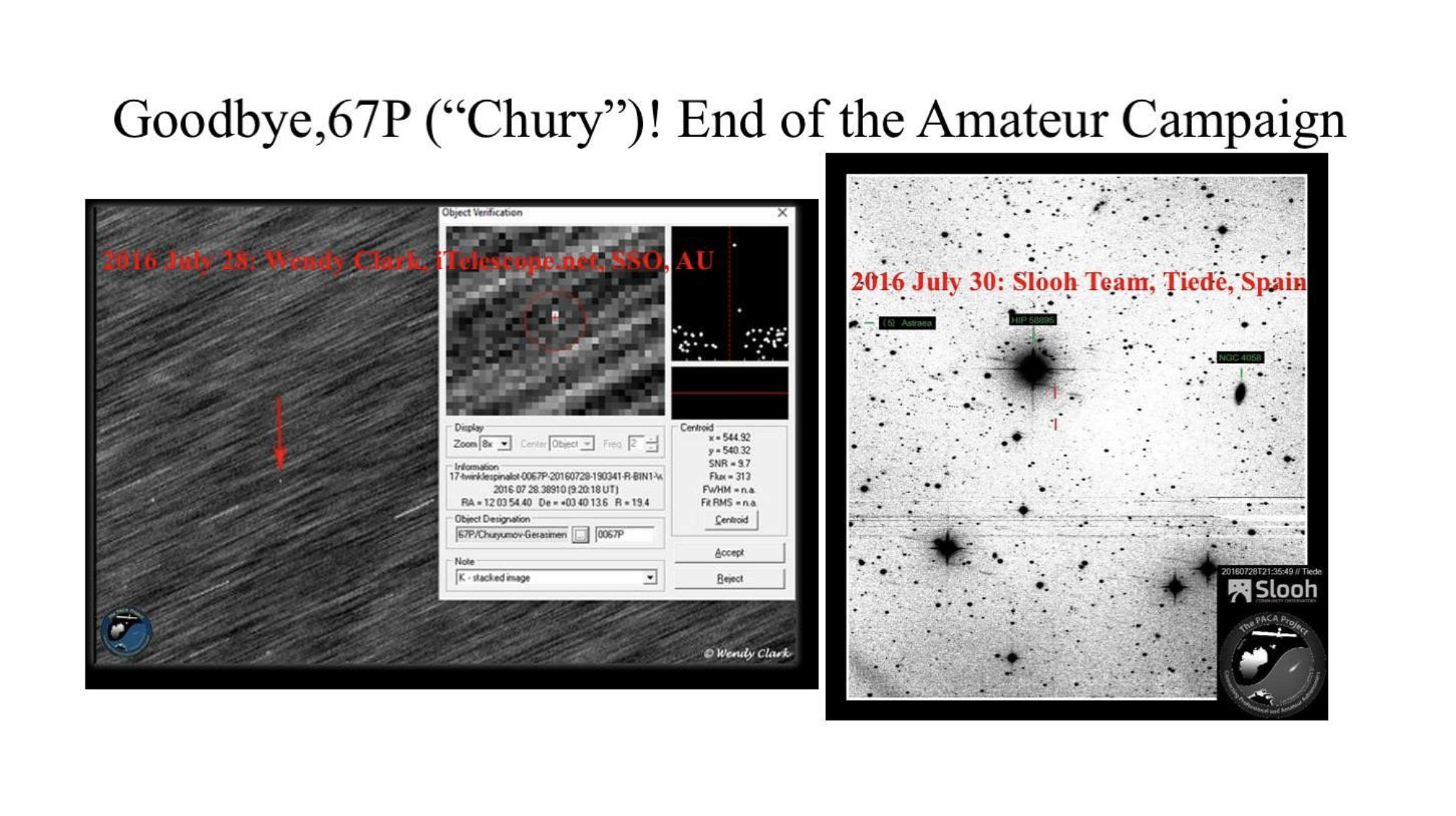}
\caption{Example of Flickr Upload - \track{Image representing a significant milestone in the mission - here the last observations on 2016 July 28 and 30.} Credit: Wendy Clark/Slooh}
\label{fig:goodbye}
\end{figure}

In addition to the submission of FITS data, members uploaded JPEG images to FLICKR and Facebook.  The PACA67/P(Churyumov-Gerasimenko) FLICKR\footnote{\url{https://tinyurl.com/Paca-67P-Flickr}} group has 272 ground-based images (1 July 2020), uploaded by 47 observers.  Of these, 36 uploaded $\leq$5, 9 between 6 and 25, and the remaining two, 36 and 56.  The majority (77\%) also included scientific analysis, primarily photometric measurements (Figure \ref{fig:tonyflickr}), but also morphology (Figure \ref{fig:ebflickr}) and screenshots from Astrometrica\footnote{\url{http://www.astrometrica.at/}}. Images at key points in the comet's orbit, or significant milestones in the mission, were often uploaded (Figure \ref{fig:goodbye}).  Some members also processed data from the mission instruments.

It is much more difficult to catalogue the uploads to Facebook, as the discussions and uploads relate not only to science data, but also to the mission more generally, and social and conference elements too.  Facebook does not lend itself to effective cataloguing and archiving of content.


\
\
\subsection{Potential Uses for the Amateur Data Set}

Amateur data can be used for astrometry, photometry and morphology.  Astrometry measures the comet's position, which allows study of changes due to non-gravitational forces caused by comet activity.  Characterisation of comet orbits is important for ensuring effective in-situ measurements, \track{ for predicting possible stellar occultations,} and also for monitoring any potential hazards for Earth. 
Photometric studies allow the measurement of total brightness, which allows monitoring of dust and gas production rates, and how they vary through the orbital/rotational cycles. 
Coma morphology, monitoring outbursts and jets from the nucleus, also gives insights into rotation and pole orientation. Such measurements can be compared with in-situ data to verify correlations between large-scale and local structures that could allow interpretation of events in comets not visited by spacecraft.

Photometry can be performed automatically using different apertures to correspond with different scales at the comet (with \track{pixel scale, and therefore aperture radius,} calculated automatically by querying the \track{HORIZONS\footnote{\url{https://ssd.jpl.nasa.gov/horizons.cgi}} database for comet distance at each observation time)}. Differential photometry techniques rely on comparisons with stars in the same frame as the comet. For amateur data there are two potential challenges to this approach: the robustness of calibration (particularly flat fielding) which could result in inconsistencies across the frame; and, knowledge of the filter and CCD response is required to ensure colour match to catalogue objects.  The A\textit{f}$\rho$ parameter can also be calculated as a way of comparing results across different telescope apertures and systems.  This is already done under the CARA project. 

For morphological study the challenge is obtaining sufficient resolution and the use of the most appropriate specialist filters (e.g., CN) which are not generally used by amateurs.  Larger amateur telescopes, and the public and schools access telescopes, such as Slooh and Faulkes telescopes, are capable of discerning fine transient features. Where there are multiple frames on one night it is possible to co-add/stack images to improve resolution and signal-to-noise ratios.

\section{Surveys}
\subsection{Registered \emph {Rosetta} Campaign Observers} \label{regsurvey}
To improve the robustness of the metadata, and understand the pre-processing of submitted data, each amateur observer was contacted directly where possible.  Feedback was sought on their experience of the campaign and its processes, and suggestions for future campaigns.  Of particular interest were the reasons why such a small percentage of those who signed up to the campaign actually submitted data.  The responses were gathered through a Google Forms survey (Appendix \ref{Annex1}) which was sent to all those who signed up for the campaign and who had previously agreed to be contacted (to meet data protection regulations).  

Thirty participants completed the survey, of whom 20 (out of 26) were observers who had submitted FITS data.  This unfortunately meant that the survey produced little useful data on why observations were not made or submitted, but it was possible to gather some information from responses to initial emails. Some people signed up for the campaign as they were interested in the mission and wanted to be kept informed, so there was never an intention to submit data.  Others suffered from poor observing conditions: weather and observability (the comet was often poorly placed and visible only during the early hours of the morning).  For some, they could not meet the requirement for submitting FITS files, having used methods of capture such as DSLRs, although some of these images were uploaded to Facebook or Flickr sites.  Some observers struggled with the technical requirements including upload.

The main results of the survey are:
\begin{itemize}
    \item Observers heard about the campaign from a wide range of sources - official website, group websites (forums, Facebook, societies), email groups, at conferences, articles in the astronomy and general press, personal recommendation (particularly Padma Yanamandra-Fisher), and inspiration from members of the \emph {Rosetta} team giving talks to local astronomical societies.
    \item The reasons for sign-up were related to wanting to be part of the \emph{Rosetta} mission and to contribute to the scientific study of comets.  
    \item While many were experienced observers, who had engaged in campaigns before (including some involved in \emph{Halley Watch}), some were new to scientific observing and were looking to enhance their skills and enjoyment.  
    \item Over half (59\%) of observations were made primarily for the campaign, 26\% primarily for personal use, with the remainder being mixed use (including submitting to other data collection organisations such as the BAA and COBS, and to forums and magazines).
\end{itemize}

\begin{figure}[ht]
\includegraphics[width=15cm]{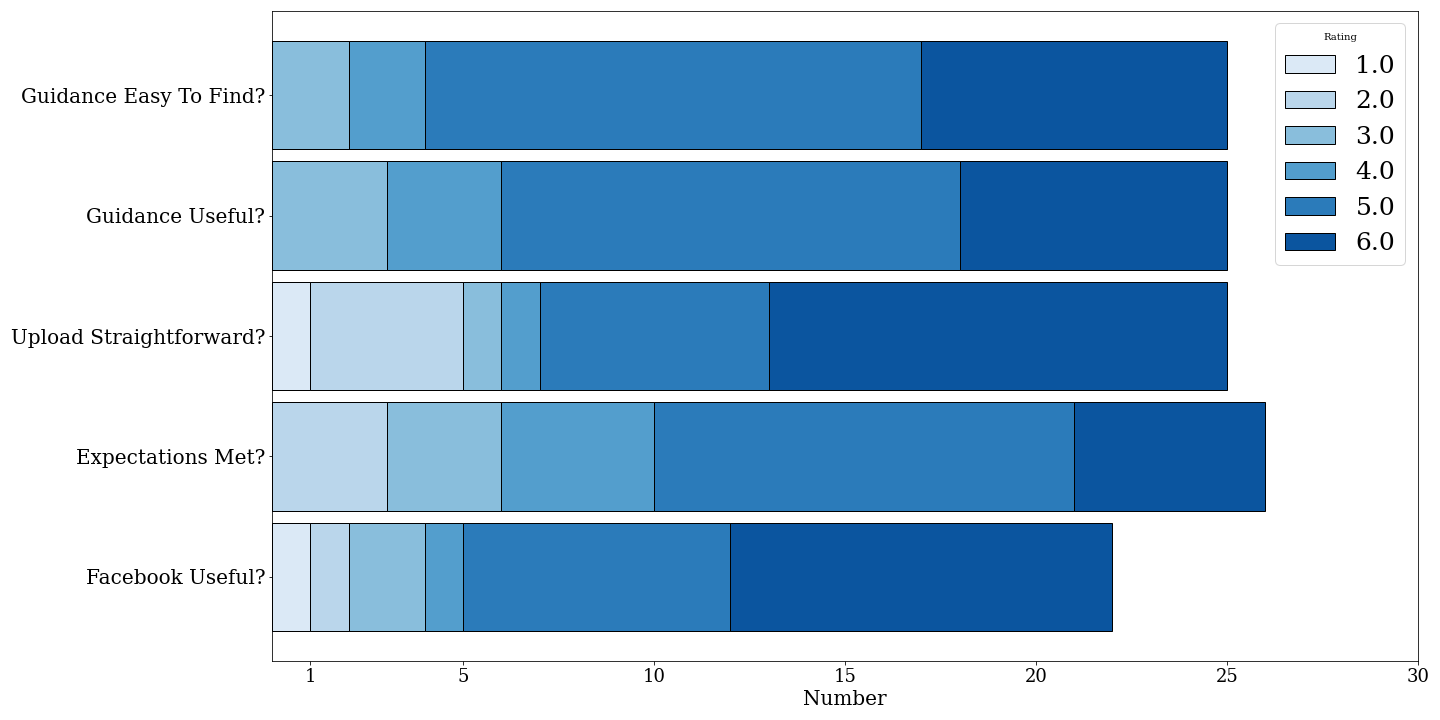}
\centering
\caption{Survey Results from Campaign Participants (Rating - 1:Low, 6:High)}
\label{fig:survey-obs}
\end{figure}

Generally, observers were happy with the guidance provided (Figure \ref{fig:survey-obs}), although some commented that publicity for the guidance could have been better.  
It became clear later in the process that some of the terminology in the guidance was interpreted differently between amateurs and professionals \track{(e.g., professionals refer to the process of applying calibration frames as 'reduction', amateurs refer to it as 'calibration', and to an amateur using Astrometrica 'reduction' is analysing the data using catalogue matching and producing measurements of position and brightness)}\footnote{\url{http://tiny.cc/TAngelSeggauPresentation}} 

For those who were members, most found the Facebook page a very useful source for advice and discussion. The group was a closed group, and this limited wider sharing of images and engagement.  Not all observers got the advice they needed, and not all of the guidance was implemented by all observers.  

There was a wide variation in both software used, and workflows.  Just over half (56\%) of observers said they submitted their observations as they made them, with 44\% submitting as a block at the end of the campaign.  Some observers found the upload process difficult (Figure \ref{fig:survey-obs}).  Additionally, it was suggested that FTP was an unsophisticated approach, and the need to manually rename, in some cases thousands of, files was onerous.

Observers suggested it would be useful to have verification processes in place at the start of a campaign to ensure compliance with FITS header requirements, and highlight any quality or compliance issues for timely resolution.

Tools for determining optimum observations (e.g., exposure times, number of frames, filters) based on each observer's equipment, location, and mount characteristics, would be welcomed.  For less experienced users, more detailed guidance (including walk-through and video guides) would be helpful.

Effective communication is critical to an effective campaign.  The survey results for communication methods show that most amateur astronomers can be traditional in their preferences for modes of communication, and many do not use social media.  The preference for email lists was common to almost all (90\%) respondents. 

A clearer understanding at the outset of the use to which the observations were to be put would have helped observers make the most useful observations.  Most contributors would welcome more information on the progress of the campaign, the analysis and results. 

A very encouraging finding was that all observers said they enjoyed being part of the campaign and were likely or very likely to participate in further campaigns.

\subsection{Amateur Astronomy Community}

A more general survey of the amateur astronomer community was also undertaken (Appendix \ref{Annex2}).  This was to gauge knowledge of the original campaign and determine what might encourage greater participation in future campaigns.  This survey was widely disseminated through societies such as the BAA, astronomy forums, comet mailing lists, Facebook (including the PACA page), Twitter and via the Royal Astronomical Society's Specialist Discussion meeting on comets in December 2019.  

Forty-four people responded, from 8 different countries (although 72\% were from the UK, reflecting the distribution methods).  Only 2 had submitted data for the campaign.  Fifty-five percent had heard of the \emph{Rosetta} campaign, having heard from a range of sources.  The main sources cited were: forums (5), BAA (4), PACA Facebook group, magazines, professional conferences/mailing, personal contact (2) and web and mailing list (all 2 each).  The survey asked a general question about where observers got their information on comets.  (This was designed to capture data on sources for publicising future campaigns.) Again there was a wide range: specialist mailing lists (e.g., comets-ml), forums, newsletters, national associations (such as BAA, Society for Popular astronomy (SPA), American Astronomical Society (AAS), Astronomy Ireland), local societies, specialist comet websites (MPC, JPL Horizons, COBS, CARA, personal websites of specialist comet observers, general astronomy websites (Astronomy Picture of the Day (APOD), weekly/monthly skyguides), social media groups (Comet Watch, PACA, on Facebook), magazines, news organisations,  remote telescope operators (e.g., iTelescope), YouTube channels, planetarium software and word of mouth.

For future campaign communication there was a clear split over social media, with 47\% not wishing to use social media.  There was a strong preference for a dedicated website and/or forum to host all the information for the campaign and allow discussion, supplemented by a mailing list and regular newsletters.

On guidance, respondents felt that availability, consistency, and detail were important.  Guides should include details of comet observability based on location, charts on how to find the comet, best equipment to use, and observing techniques. The level of detail should be tailored to different observing cohorts (general public, schools, general observers, specialist and experienced comet observers). The science observations' guidance should cover the purpose of the observations, ensuring accurate timing, requirements for FITS headers, and the provision of calibration files or evidence of appropriate pipeline processes.  All terminology should be clearly explained to ensure consistency and avoid confusion.  

Tools could be developed for planning observations (e.g., to calculate optimum exposure times based on equipment, the movement of the comet, and the purpose).  The upload process should be simple, incorporate a compliance check for FITS header information, and automatically generate filenames with the naming convention.  It should be made easy to provide brief context data \track{ e.g., weather conditions, any issues with the observing}.

Where initial analysis was to be undertaken by observers (particularly for novice observers and schools) detailed walk-through guides, and video tutorials should be prepared.  

Two particularly interesting ideas were: work with mobile app providers (e.g., developers of planetarium tools such as SkySafari, Stellarium) to provide both publicity and guidance as part of the app (e.g., inclusion in \lq Tonight's best' recommendations, alerts for observing opportunities); and secondly set up a mentoring scheme to provide detailed help and guidance.  A dedicated forum would help the community share knowledge, experience and to keep enthusiasm - as well as disseminating and showcasing results.

To encourage involvement in future campaigns, respondents said that a clear statement of the scientific value amateurs can add to campaigns was the first priority, related to campaign aims, objectives and outcomes. All possible communication channels should be used for initial and ongoing communication -- one size does not fit all.  Ideally, a `buzz' should be created around the campaign, in the mainstream media if possible, magazine, astronomy societies, videos, web and social media.  Outreach and schools' events would also bring the campaign to new audiences.  The campaign could set up some student projects, which could report through teacher and learning networks -- possibly linked to societies or academics.

Competitions could be organised to generate wider interest (e.g., first sighting of comet, best images, best sketch, best with a smart phone).  These images (rather than science data) in a gallery could be a rich source for publicity and illustration purposes. 

The outputs from the campaign, in terms of both scientific output (posters, conference presentations, research papers, press releases) with amateurs as co-authors or cited data submitters, and a data archive for future use, should be regularly reported.

\subsection{Faulkes Telescope Project Comet 46P/Wirtanen Schools Campaign}
A campaign\footnote{\url{http://resources.faulkes-telescope.com/course/view.php?id=150}} of observations of comet 46P during its close approach to Earth in 2018/9 was set up to test the feasibility of running a campaign aimed at schools (through the Faulkes Telescope Project/Las Cumbres Observatory \citep{Brown2013}) and to test processes and guidance.

The campaign included developing: background materials on comets; details on observing 46P including finder charts; walk-through guides for setting up observations; details of observations required; and, detailed guides for astrometric, photometric and morphological analysis.  The project also provided some hands-on support for teachers.  

In total 2,638 observations were made during the period 1 June 2018 to 30 April 2019 (not all directly from the campaign).  

To assess the effectiveness, and learn lessons, a third survey (Appendix \ref{Annex3}) was undertaken of those UK schools who had participated. All three submitted their feedback \track{- see Acknowledgements}.
 
Schools said they chose to participate to inspire their pupils in science and astronomy, using real research.  They heard about the campaign through an astronomy forum and the Faulkes Telescope Project mailing list.  Sixty-two pupils participated, 30 in primary (state school) and 32 in secondary (private schools).  There was a mix of whole-class participation, astronomy clubs and individual pupils.  All pupils were involved in scheduling observations on the LCO telescope network, processing and analysing the data.  All schools said their pupils enjoyed being part of the campaign, and that the enthusiasm was maintained through the three months of the campaign. 

The guidance was considered useful, but more-detailed guidance on processing would have been helpful, perhaps in the form of videos.  All felt that a forum for discussion with other educators would be a useful addition for future campaigns.

Those leading the work in their school said the project was engaging, it allowed them to share their love of astronomy and engage their pupils (and their parents) in comet observations.  It provided a catalyst for developing after-school astronomy observing sessions, and for science activities around solar observing (during school day). 

The educational value was considered to be broad.  One school was a girls school, and this project inspired them to be more involved in physics and science.  Others said the combination of astronomy, physics, chemistry, maths, geography, and planning (including dealing with different time zones) made for a rich educational experience. All would like to be involved in future campaigns.
\\

\section{Discussion}

Pro-Am campaigns have demonstrated that amateurs can add value, particularly by providing better temporal coverage.  What can be learnt from the effectiveness of these campaigns, and the \emph {Rosetta} campaign in particular to inform future \track{amateur} campaigns?

Older campaigns had greater logistical challenges due to the lack of modern communication methods.  More modern ones have potentially better communications and better equipped amateurs \track{. Ensuring adequate mission/campaign resources to actively manage the planning, implementation and follow up is always a challenge}.  Process and cost efficiency is essential, and this means effective planning, clear guidance, tools for observers, effective initial quality control, and realistic and robust plans for collecting and archiving the data.

\subsection{Campaign Objectives}

It is important to be clear about the goals of the \track{amateur elements} of a campaign.  Obtaining high-quality science data is usually the primary goal, to allow long-term analysis and short-term alerting of the professionals to significant changes in the comet.  But to look only for the best scientific data risks missing many other potential campaign benefits, for example:
\begin{itemize}
\item increasing science capital by raising awareness of comets, and astronomy, for the general public.  This is particularly important for campaigns in support of space missions, with their associated large publicly-funded costs.
\item deepening the skills, interest and knowledge of amateur observers - adding a new dimension to their \lq hobby' (although for many it is a very serious affair).  
\item involving schools can increase the interest in astronomy, science and other related disciplines.  It can also widen horizons on career choices. Observing and studying comets can be a fun vehicle for teaching a wide range of subjects -- as the survey from the 46P noted, students practiced their maths, geography, physics, biology, chemistry, planning, cooperation and analysis skills.  They gained insights into the way real research is undertaken, including the challenges of equipment failure, software problems, and weather.
\end{itemize}

\subsection{Data Collection}

What, where, how and when data should be submitted can be difficult to optimise.  Amateurs do not have to submit their data, and are less likely to do so if the requirements are perceived to be too onerous, \track{but without compliance with appropriate standards submitted  data can be almost useless.}
\subsubsection{What to Collect?}
If the campaign is looking to analyse morphological changes in the comet over a long period then multiple images stacked, repeated over multiple nights, will give good SNR to allow faint detail to be teased out. If looking to constrain the start of outbursts, then the submission of individual accurately timed, high-cadence images is important (even though these might be low SNR). Larger aperture telescopes will provide the best resolution, although tracking is more of a constraint. For large-scale features such as large comae and dust and gas tails, smaller telescopes with wider fields of view will be most suited.  Longer exposures are also possible before star or comet trailing becomes an issue.

For some purposes it may be useful to receive the results of analysis, rather than raw data.   An example would be astrometry measurements from standard software packages such as the widely-used Astrometrica. The CARA project provided observers with its own developed software\footnote{\url{http://cara.uai.it/soft_list}} to measure A\textit{f}$\rho$ in a consistent way, and the results were collected and collated, rather than raw data.  For the \emph{Deep Impact} mission photometric measurements were requested, with raw data FITS files only submitted later.  The \emph{ExoClock}\footnote{\url{https://www.exoclock.space/}} project also provides software and an agreed methodology (for measuring exoplanet \track{transit} lightcurves), as does the Lunar Impact Flash\footnote{\url{https://www.nasa.gov/centers/marshall/news/lunar/observing_schedule.html}} project (for detecting and measuring lunar meteor strikes).   Robust guidelines, good tutorials, and, ideally, provided software are key requirements for making these types of submission useful.

The \emph{GAIA} alert\footnote{\url{http://gsaweb.ast.cam.ac.uk/alerts/home}} follow-up project takes a slightly different approach with observers asked to do some initial data analysis (with Astrometry.net\footnote{\url{http://astrometry.net/}} and Sextractor\footnote{\url{https://sextractor.readthedocs.io/en/latest/}}) before uploading the results to a calibration server\footnote{\url{http://gsaweb.ast.cam.ac.uk/followup}}.  This server calculates magnitude, without needing knowledge of filter used, and populates a live lightcurve for each \emph{GAIA} alert object with data points credited to the observer.

In presenting science results, particularly when engaging with the media, and engaging schools, it is very helpful to have good-quality colour images of the comet.  Producing colour images from multiple science filters is tricky - not least because the comet may move significantly between the images taken in subsequent filters.  So a single (or better, stacked) colour image taken with a standard digital camera or a one-shot-colour astronomy camera can really add value for publicity and public engagement purposes.  For these, precise timing is not important, nor many details of the capture and processing.  This opens up the campaign to a much broader group of astronomers and even the general public (as demonstrated by the multiple images of C/2020 F3 NEOWISE posted on social media and websites).

Clear guidance on what files (images, calibration files), what format (FITS and what FITS headers, JPEGS, other pictures) and what processing can or cannot be done is critical.  \track{This must be available before the start of the campaign, and stressed during the campaign - allowing observers to decide whether they are prepared to spend the time and effort needed for science observations.}    Science data should be unprocessed, \track{and to be useful must be accompanied by specific metadata (e.g., accurate timing, exposure length, filter, sensor details).  Other metadata (e.g., context data) are useful but not essential.}  For publicity or educational purposes, JPEGs are acceptable, and enhancement techniques are useful, while details such as precise timing are less critical.  \track{Given the different levels of rigour needed, it would be advisable to set up different, clearly differentiated, channels for submission.  The process for pictures could be much simpler.}

\subsubsection{Where to Upload?}
The  decision on where to upload and how to archive is difficult, particularly for smaller campaigns.  For \emph{Rosetta}, the ESA's PSA archive was planned as the repository.  Late set-up, a lack of clear guidance to users, and confusion over necessary filename conventions, meant that the data and observations were split between uploading via FTP to ESA storage, a Dropbox facility, FLICKR and Facebook pages. 

While the FLICKR site currently houses a very useful archive of images, the absence of cataloguing makes it difficult and time-consuming to locate any specific observations.  For Facebook it is even more difficult, and now that the group (which was members-only) has been archived the images are not publicly available.  Both FLICKR and Facebook rely on private companies for existence, and their future cannot be guaranteed.

ESA's PSA archive standards are stringent to ensure long-term accessibility and compatibility.  The time and cost of converting all the amateur data to a consistent format is unlikely to be a priority for ESA or another agency.  For \emph{Halley Watch}, all data were initially held in hardcopy, before being digitised on CD, and made available online at NASA's PDS: Small Bodies Node.  The \emph{Rosetta} archive could similarly be stored but not converted into a future-proof format or catalogued in detail.  The filename convention adopted for upload to the ESA PSA FTP site (Observation date\_UTC Time\_ Object\_Filter\_Exposure in seconds\_Observer initials.FITS) is good and would be sufficient for any future researchers to at least identify date of observation, filter and observer.  With an index (of observer and their equipment and location) this would allow for a quick filtering of observations for any purpose and this method may be appropriate for future campaigns too. The challenge is to decide who will provide the storage and the accessibility.  It is also worth noting that even conversion of files to a standard naming convention appeared to be a barrier to participation to some observers (given the large number of observations they made), with most of the files uploaded via Dropbox eventually being renamed by a JPL intern.  

\subsubsection{How and When to Upload?}

The key is simplicity but robustness. With modern large-chip, high-resolution images, file sizes are large.  If multiple observations are made over a night then the amount of data needing to be uploaded becomes multiple Gb.  In some parts of the world this is not an issue, but in remote locations internet speeds are slow and connection costly. A way of compressing data for upload is important.  Ideally a web-based interface (rather than an FTP or similar system) needs to be provided with a zipping tool to save bandwidth built in. Quality control should be built in - verifying FITS headers, and highlighting non-compliance early enough for corrections to be made.  The system should generate consistent filenames to be used as an access tool.    A log should be kept of all observations uploaded, by observer, with context and contact details, and this should form a key part of the archive.  Ideally, observations should be uploaded as soon after they are made as possible, along with a short covering narrative. 

\subsubsection{A Long-term Collaborative Comet Campaign Website and Archive?}

The \emph{Halley Watch} project has demonstrated that having access to a digital archive can result in extra analysis long after the campaign - data analysis tools and techniques improve over time.
There are currently a number of organisations who take either observation files or observational data\track{ from amateurs }(eg BAA takes JPEG images, COBS and MPC take astrometric and photometric results).  For the latter, consistency of measurement technique (particularly apertures used) is challenging, and this constrains the robustness of the data. 

If a longer-term, more generic solution is considered \track{ (potentially including professional data too),} there are many practical questions to be addressed: who should host the website and upload facilities, who should store comet data, how would it be quality-controlled, how long should it be kept, with what access, and how could the management and support costs funded. 

In the short term, in Europe, the Europlanet VESPA\footnote{\url{http://www.europlanet-vespa.eu/}} programme may be able to help.  The Planetary Virtual Observatory and Laboratory (PVOL)\footnote{\url{http://pvol2.ehu.eus/pvol2/}} database \citep{Hueso2018} is an example of a VESPA-funded project.  It makes available planetary images taken by amateurs across the world, with consistent meta-data. Unfortunately the Europlanet programme is funded in short-term blocks by the \track{European Commission}, so its long-term future cannot be guaranteed.     

\subsection{Effective Communication}

Modern communication methods should make effective communication much easier than earlier campaigns - although the existence of multiple channels adds complexity. There is a split between observers who use social media and those who do not, and this needs to be factored in. A website to hold all the guidance and tools (including upload), live updates, feedback to observers, and a discussion forum is the foundation. There are established interactive mailing lists with a wide membership such as Comets-ml.  There are also a few core comet and Pro-Am Facebook groups. These should be used.  Traditional print media (magazines, newspapers) may be reducing in number, but still have a place, along with their digital arms, for getting messages out to observers and the general public.  Local and national societies provide good access to traditional (and often highly-skilled) observers, and internet forums provide access to active communities too. The personal touch should not be forgotten - some observers in the \emph {Rosetta} campaign became involved after a talk at their astronomy society from the mission scientist Matt Taylor. Two schools were involved in the 46P campaign due to personal contact with the organiser. Personal requests from Padma Yanamandra-Fisher also led to experienced observers joining the campaign. 

Core messages and information and guidance need to be consistent however they are communicated, but modified for specific audiences.  Regular communication, during both the data gathering and subsequent analysis stages, is key to keeping observers engaged and enthusiastic, as is recognition and credit in publications.  

\section{Conclusions}

\subsection{Campaign Summary}
The \track{comet} 67P amateur campaign certainly created interest in the \emph {Rosetta} mission: 10,432 observations were submitted by 26 observers/groups, covering 284 dates.  This compares with 17,352 observations over 463 dates by professionals. There are 58 days during the main observing period (2013-04-17 to 2016-04-30), and 15 in the 3-month period around perihelion in August 2015 when amateur but no professional data are available. So amateurs have added significantly to the observational coverage. There is good longitudinal coverage (Figure \ref{fig:obslocation}), and wide scale variations (Table \ref{tab:obs}).

\subsection{Surveys Summary}

In total 77 people responded to the surveys:
\begin{itemize}
\item Observers and the wider astronomy community felt clarity of purpose and guidance, and regular communication were the most important elements of a campaign.  Data submission should be made straightforward, with tools to ensure compliance with standards. There was clearly room for improvement in both of these areas in the \emph {Rosetta} campaign.
\item Useful metadata were collected as part of the survey to supplement/correct data from FITS headers. Having these data submitted in a consistent format with the observations would have been better, and should be implemented for the future.
\item Observers really enjoyed being part of the 67P campaign and would wish to be involved in future.
\item Educators said the schools campaign had wide educational benefits, as well as being enjoyable and inspiring for pupils, staff and parents.
 \end{itemize} 

\subsection{Elements of an Ideal Campaign}
The survey results, along with analysis of previous and current amateur observing campaigns, have informed the following suggested elements of an ideal campaign.  \track{While these are framed in terms of a comet campaign, many of the principles and actions would also be applicable to other non-comet campaigns.}

\begin{enumerate}

\item Agree clear aims and objectives for both science outcomes and wider benefits.
\item Agree the observations and other data/images to be collected. 
\item Be realistic, given the resources available to run the campaign, and the uncertainty of comet brightness.
\item Prepare well in advance \track{ and learn from other campaigns (re-using material where appropriate).  Involve the amateur community, and the professionals who will use the data, in the planning.}
\track{\item Build in a test phase well before the campaign is due to start.  This should include sample observations, by a range of observers, to test the processes, systems and guidance.  The feedback from both observers and researchers will allow refinement and streamlining (e.g., minimum metadata required, ease of upload, clarity of guidance), so that the actual campaign data are not compromised. It will also establish a set of experienced super-users who may support the community and act as mentors.}
\item Set up a campaign website to be the information hub: repository for guidance (at various levels), tools, feedback, forum for discussion, and uploading data.  (In the longer term this could become be an overarching website covering many campaigns.)
\item Carefully consider the launch elements so that the momentum can be maintained.  This may mean launching different elements, for different cohorts, at different times.
\item Use a wide variety of communication routes: press releases, astronomy press, societies of all sizes, mailing lists, forums and social media.  But keep everything consistent and try to draft once then disseminate, not cover everything individually.  Create a buzz around the campaign by running competitions (e.g., first sighting, first image with different size telescopes, art competitions).  Contact the main software providers, particularly app developers, and engage them to include in bulletins, highlights lists and observing alerts. (This will be dependent on the expected brightness and observability of the comet.)
\item Provide tools for observing: guides to position, optimum observing and exposure times; ideally these should be tailored for each observer's location and equipment (as with \emph{Exoclock}\footnote{\url{https://www.exoworldsspies.com/en/observers/}} project).  There should be more general information for novice observers and more technical for experienced observers, including details of ideal filter specification.  Develop tools and guidance to allow DSLR users to submit scientific observations, if the comet is expected to be bright enough,  e.g., to ensure proper timing, as this will open up the campaign to many more observers (see the deluge of DSLR images of comet C/2020 F3 NEOWISE) and be particularly useful where viewing conditions are difficult due to low altitude, and/or short observing windows. 
\item Where practical the guidance should include multi-media, e.g., short video tutorials and walk-through guides (particularly for the educational aspects).  Consider setting up a mentoring scheme using experienced amateurs to guide other amateurs and schools.
\item Use the website forum to allow real-time discussion and provision of advice.  Encourage participants to share their observing experiences as well as data.  For educators, encourage them to share how they are using the campaign in classes and activities.
\item For upload, make it easy, ideally with compression to save bandwidth. Keep it to one location, with timely verification of data submitted via FITS tool, plus a short narrative for context information. Use a naming convention which can be used to search for data, but automate file naming on collection rather than introducing additional complications for observers.  Remember that analysis techniques will improve over time so having an archive will be a legacy for future astronomers.
\item Ideally, following upload there should be a pipeline process to quickly measure magnitude and position (if the observer has not already reported to MPC).  The magnitude should be logged on a real-time lightcurve, with data points credited to observers (like \emph{GAIA}).  This should be on the front page of the campaign website.
\item Provision of regular updates on what is happening with the campaign and what research is being undertaken is key to keeping observers engaged and valued for both the current and future campaigns.
\item Recognise all submissions as adding value (e.g., produce certificates of contribution to campaign).
\item Make the final data set freely available, and accessible, using the FAIR principles (Findability, Accessibility, Interoperability, and Reusable) \citep{Wilkinson2016}. 
\item Undertake a post-campaign evaluation to learn and disseminate lessons for future campaigns.
\item Celebrate success.
\end{enumerate}

Comet 67P returns to perihelion in November 2021, and is favourably placed for observation from ground-based telescopes.  This apparition will provide an excellent opportunity to test the observing campaign principles and good practice set out in this paper.  The resultant data can be analysed alongside the earlier campaign data to learn more about the evolution of this favourite comet.
\\

 \acknowledgments
 {We would like to acknowledge the inspiration of Mike A'Hearn.  We hope we can continue his legacy by supporting Pro-Am comet observing campaigns in the future.
 
 We are grateful to Dr Padma Yanamandra-Fisher for providing insight into the campaign, and support and encouragement to the amateur observers during the campaign. Tony Angel and Wendy Clark have freely shared their practical experiences of participation in the campaign, and their experiences of observing comets more generally.  We thank them. 
 
Mary Abgarian provided access to the amateur data held by JPL and Nicolas Ligier kindly provided data on the observations from the professional campaign.  

The Faulkes Telescope Project provided access to the \track{Las Cumbres Observatory }telescope network for the 46P schools' campaign.  
The main contributors to the campaign were St Mary's Catholic Primary School, Bridgend; RGS Dodderhill School, Droitwich Spa; and Marlborough College, Marlborough.  We are grateful for the enthusiasm of their pupils and staff, particularly Ben Wooding, John McGrath and Gavin James, and hope we have inspired some future comet scientists.  

We would like to thank Elizabeth Warner, University of Maryland, for providing information on the practical administration of previous amateur campaigns and on gathering observer feedback.  These were an invaluable starting point in the design of the surveys for this work.

\track{ We thank the reviewers for their helpful and constructive comments. } 
 
 Last, but certainly not least, we are grateful for the time, skill and enthusiasm of the observers who have submitted data and images. In addition to those who submitted FITS data (shown in Table 1), the following observers submitted images to the PACA Flickr group:  P Yanamandra-Fisher, V Agnihotri, B Backman, A Baransky, J G Bosch, D Buczynski, M Bunnell, P Camilleri, K Churyumov, G Conzo,  P Cox, D Eagle, G Fagiolo,  C Feliciano, F Garcia, J Gonzalez,  N James, M Kardasis, R Kaufman, S Kunihiro, D Lovro, A Maury, G Masi,  R Miles, E Morales, R Naves, T Noel, A Novichonok, A Oksanen, D Peach, T Prystavski, D Romeu, K Sugawara,  K Takeshita, J Tillbrook, A Tough, J Tuten, S White, A Yoneda.}

\software {\track{The data analyses were undertaken using}
Astropy \citep{astropy2013, astropy2018} and Numpy \citep{Harris2020}. \track{ The plots were generated with} Matplotlib \citep{Caswell2020}.  \track{The pixel scales were extracted using astrometry.net \footnote{\url{http://astrometry.net/}}. The Google Map was produced using jupyter-gmaps\footnote{\url{https://github.com/pbugnion/gmaps}}.}}
\newpage
\appendix
\section{Survey Questions: Registered \emph {Rosetta} Campaign Observers}

This questionnaire seeks your experience of the Amateur Observing Campaign in support of ESA's \emph {Rosetta} mission to comet 67P. It also invites you to submit details of any observations, and your opinions on how future campaigns could build on the \emph {Rosetta} campaign. This is part of a PhD research project being undertaken by Helen Usher at the Open University, UK, under the supervision of Dr Colin Snodgrass.
Personal details provided will only be used for the purposes of this research (at Open University). No personal details will be released, except to give you credit for the observations you made (and you will be informed beforehand).
If you have any questions on this research please feel free to contact Helen Usher directly - helen.usher@open.ac.uk
\begin{enumerate}
 \item What sources do you use for information on comet observing (please give as many details as possible eg which websites, magazines) ?
 \item Membership of Astronomy Groups
 \item How did you hear about the amateur campaign?
 \item What sources do you use for information on comet observing (please give as many details as possible eg which websites, magazines) ?
 \item Why did you sign up?
 \item Are you an observer, or someone just interested in the campaign?
 \item Did you make observations?
 \item If you didn't make observations, could you briefly explain why not
 \item Were your observations primarily for personal, primarily to submit to the campaign?
 \item Dates of observations
 \item What guidance did you refer to before making observations?
 \item Where did you access the guidance? (JPL/ESA/Facebook/Other)
 \item How easy was it to find the guidance? (1-6)
 \item How clear and useful was this guidance? (1-6)
 \item What factors led you to give the score above?
 \item Did you use a remote shared facility?(iTelescope/Slooh/FT/Other)
 \item Did you use your own equipment?
Location of telescope, description, aperture, focal length, camera type, make and model, make and type of filters used.
 \item What software (if any) did you use for acquisition?
 \item Could you provide details of your acquisition workflow?
 \item If you calibrated your images before submission what software did you use?
 \item What was your calibration workflow?
 \item What software did you use for any processing? 
 \item What was your processing workflow?
 \item Did you submit your observations?(Y/N)
 \item If you did not submit could you tell us why not?
 \item When did you submit observations ?(As I made them/All at once at the end of campaign)
 \item Did you submit to (ESA FTP/via P Yanamandra-Fisher/Facebook/Flickr)
 \item Did you submit (Calibrated FITS/RAW FITS/JPEGS/Calibration files/Context info)
 \item How straightforward did you find the upload process? (1-6)
 \item What factors led you to give the score above?
 \item If you uploaded FITS files did you ensure the FITS headers contained all the required observation data?
 \item How could we help you to easily provide these FITS header data in future? (accurate FITS header data makes analysis much easier and more robust)
 \item When you registered what were you expecting (including support, guidance, on-going communication)?
 \item What sources did you use to obtain the information and guidance on the campaign (please be as explicit as possible)?
 \item How well were your expectations and needs met? (1-6)
 \item What factors led you to give the score above?
 \item Did you join the Facebook Group?(Y/N)
 \item If no, could you give details of why not, and what you would have preferred instead?
 \item If yes, how useful did you find the Facebook group (1-6)?
 \item What factors led you to give the score above?
 \item Did you post images and/or comments?(Y/N)
 \item If there was a future similar campaign (eg 67P at next apparition) would you be likely to participate? (Definitely/Probably/Possibly/No)
 \item Was there any information (or were there any tools) which would have made observation and upload easier for you?
 \item What are your preferred methods of communication? (Email mailing list/Website/Social Media/Dedicated forum/Dedicated group message board/Regular online newsletters/Magazines/Microsoft teams or similar/Other)
 \item Is there anything you feel should be done differently for future campaigns?
 \item Are you aware of any other professional-amateur collaborations and observing campaigns which are particularly effective, and from which we might draw good practice lessons?
 \item How should other observers be encouraged to be part of future campaigns?
 \item Any other comments/suggestions/complaints/kudos/answers to unasked questions? 
 \item Finally, did you have fun?
\end{enumerate}

\label{Annex1}

\section{Survey Questions: Amateur Astronomers}
The ESA \emph {Rosetta} mission to comet 67P included an amateur observing campaign.  The aim was to encourage amateurs across the world to submit observations of the comet, which could then be used to supplement professional observations.   Amateur data can add greater temporal sampling and wider fields of view. 

This questionnaire, which forms part of a PhD study by Helen Usher at the Open University,  seeks information on the effectiveness of the awareness raising methods used, and seeks views on how future observing campaigns could most effectively reach comet observers worldwide.

The personal details provided for the purposes of this research (at Open University).  No personal details will be released.

If you have any questions on this research please feel free to contact Helen Usher directly -  helen.usher@open.ac.uk

\begin{enumerate}
\item Country
\item What sources do you use for information on comet observing (please give as many details as possible eg which websites, magazines) ?
\item Membership of Astronomy Groups
\item Did you know that there was an official amateur astronomer campaign in support of the \emph {Rosetta} space mission to comet 67P?  If so, can you remember where you heard about it?
\item Did you participate in the campaign? If you participated in the campaign, have you received the more detailed survey for participants from Helen Usher? (If not, it is available here https://forms.gle/iUMeLYMu5SVguAqVA)
\item How should observers be encouraged to be part of future campaigns?
\item What publicity should be used?
\item What guidance and tools should be provided?
\item How should the guidance and tools be made available?
\item How should ongoing communication be handled? \item Any other comments?
\end{enumerate}
\label{Annex2}

\section{Survey Questions: 46P Schools' Campaign Observers}
Thank you for participating in the campaign.   We hope you enjoyed being part of it, and it provided good learning opportunities for (you and) your pupils.

This was the first time we have really attempted a comet observing campaign, but we hope to do more in the future!  We would therefore be very grateful if you could fill in this short questionnaire to let us know what was good and useful, and what could be improved.

As well as informing future FT/LCO campaigns, Helen Usher will be drawing out more general lessons as part of her PhD studies with the Open University, UK.

If you are happy for Helen to follow-up then please include your name and contact details.  The data will be kept securely and used purely for the purposes of this research.  No names will be released without prior approval.

Thank you!  Helen Usher and the FT team

\begin{enumerate}
\item Your name, role, school
\item School type (Primary/Secondary)
\item Why did you decide to observe/be part of the 46P observing campaign?
\item How did you hear about the FT campaign?
\item How many pupils involved? (age range)
\item Did you use the activities with (whole class/astronomy group/individual or selected pupils)
\item What activities did you undertake?
\item How much did your pupils enjoy being part of the campaign? (1-6)
\item What factors led you to give the score above?
\item How much did you enjoy being part of the campaign? (1-6)
\item What factors led you to give the score above?
\item What do you consider to be the educational value of the campaign?
\item How useful was the guidance? [Listed]
\item Was there any guidance missing?
\item How do you think the guidance could be improved for future campaigns (particularly any that you rated not useful)?
\item What would be your preferred method of communication with the FT campaign team for guidance etc?(FT website/FT Facebook/Twitter/Email/Discussion forum/Microsoft teams/In person/other)
\item Would it be useful to be able to discuss and share with other schools, and if so how?
\item Would you like to be involved in future campaigns?
\item How would you encourage other schools to be involved in future?
\item Any other comments?
 \end{enumerate}

\label{Annex3}

\bibliography{references.bib}{}
\bibliographystyle{aasjournal}

\end{document}